\documentclass[twocolumn]{aa}

\usepackage{graphicx} 
\usepackage{epstopdf}
\usepackage{txfonts}
\usepackage[breaklinks=true]{hyperref}
\usepackage{url}
\usepackage{siunitx}
\usepackage[bottom]{footmisc}
\usepackage{textgreek}           % needed for micrometer "um"=\textmu{m}

\newcommand{\subscript}[1]{\ensuremath{_{\textrm{#1}}}} 
\newcommand{\romannum}[1]{%
   \ifnum#1<1
     \ifnum#1=0
        o%
     \else
        -\romannumeral -#1%
     \fi
   \else
     \romannumeral #1%
\fi}
\DeclareRobustCommand{\Romannum}[1]{\MakeUppercase{\romannum{#1}}}

\begin{document}

   \title{22\,GHz Water Maser Search in 37 Nearby Galaxies}
   \subtitle{Four New Water Megamasers in Seyfert 2 and OH Maser/Absorber Galaxies}
%  Final authors list after J.Braatz, C. Henkel and A. L. Roy considered their contributions to be 
%  sufficient for acknowledgement but not co-authorship:
   \author{J. Wagner}
   \institute{Max Planck Institute for Radio Astronomy, Auf dem H\"ugel 69, Bonn, 53121 Germany \\ 
		\email{jwagner@mpifr.de}}

   \date{Received September 12, 2013 / Accepted October 21, 2013}
 
  \abstract    % \abstract{}{}{}{}{} with 5 mandatory tokens
% context: optional
   {}
% aims: mandatory
   {We report four new 22\,GHz H$_2$O water masers found in a Green Bank Telescope search toward 37 nearby objects. Our goal was to find new maser galaxies, AGN disk-masers, and objects where both hydroxyl and water maser species coexist.}
% method: mandatory
   {We observed 37 sources within 250\,Mpc that were selected by high X-ray luminosity (${L_\mathrm{X} > 10^{40}}$\,W) and high absorbing column density (${N_\mathrm{H} \gtrsim 10^{22}}$\,cm$^{-2}$). Sources included dual or triple AGN and interacting systems. We also searched objects detected in hydroxyl (OH). A catalog of 4038 known H$_2$O (non-)detections was assembled to avoid unnecessary reobservations. The final selection consisted of 16 new sources, 13 non-detections to follow up with a factor ten higher sensitivity, 10 OH masers and one deep OH absorber, of which 37 were observed.}
% results: mandatory
  {Water megamasers were detected towards the Sy~2 galaxy 2MFGC~13581, towards the 6\,GHz OH absorber NGC~4261, and towards the two 1.6\,GHz OH maser sources IRAS~17526+3253 and IRAS~20550+1656. We set upper limits on 33 non-detections. The detection rate was 25\,\% in OH galaxies and 11\,\% overall. The mean sensitivity was 4\,mJy over 24.4\,kHz (0.31\,km\,s$^{-1}$), or between $0.1\,L_\sun$ and $1.0\,L_\sun$ rms for the distances covered by the source sample. Combined with other searches, a total of 95 objects have now been searched for both OH and H$_2$O masers.}
% conclusions heading: optional
   {The maser features in 2MFGC~13581 are typical of a sub-parsec accretion disk whereas NGC~4261 likely has jet-masers in a masing torus. The NGC~4261 galaxy (3C~270; dusty torus, twin-jet) and its masers appear similar to NGC~1052 where continuum seed emission by a twin-jet supports masers in the torus. VLBI imaging is required to determine the masing regions in NGC~4261 and 2MFGC~13581. IRAS~17526+3253 has narrow $350\,L_{\sun}$ systemic masers and the tentative 5$\sigma$ detection in IRAS~20550+1656 (II~Zw~96) strongly resembles massive star formation kilomasers in NGC~2146. The latter two detections increase the number of known "dual-species" objects containing both OH and H$_2$O masers to eight. Further we found the overall dual-species detection rate (8 in 95) to be of the order of the joint probability of both species independently occurring in the same object (1\,\% lower bound). However, this needs to be verified by a more detailed analysis that accounts for the individual selection criteria of the 95 searched objects. Lastly, we see a lack of H$_2$O kilomasers in OH megamaser objects already noted by \citet{2011AA...525A..91T}. This may be due to sensitivity bias rather than astrophysical reasons.}


   \authorrunning{J. Wagner}
   \titlerunning{Four New Water Masers in Seyfert 2 and OH Maser/Absorber Galaxies}
   \maketitle

%
%_________________________________________________________________________________________________________________

\section{Introduction}

Water masers (H$_2$O; 22.23508\,GHz, ${6_{16} \rightarrow 5_{23}}$) and hydroxyl masers (OH; 1665\,MHz and 1667\,MHz) in the megamaser class have luminosities that far exceed their Galactic counterparts. They are best known as probes into some extreme environments such as merging galaxies, nuclear starbursts in molecular tori at 100\,pc scales in the case of OH, and AGN accretion disks at sub-pc scales, outflows, and the vicinity of jets in the case of H$_2$O. Water disk-masers are of particular interest and are found between an outer circumnuclear region of molecular gas that is too cold for maser excitation, and an inner atomic gas region that is too hot ($>8000$\,K) for the molecular gas phase. The kinematics and VLBI angular positions of disk- or torus-masers allow a well-constrained estimate of the central binding mass as in Mrk~273 \citep{2004AA...419..887K}, with more accurate mass estimates made with H$_2$O megamaser observations \citep{2011ApJ...727...20K}. Additionally, H$_2$O disk-masers like in UGC~3789 allow a direct measurement of angular diameter distance and the Hubble constant \citep{2013ApJ...767..154R}.

To date, over 4030 galaxies have been searched for H$_2$O masers, resulting in about 150 detections that include about 20 disk-maser galaxies (see e.g., \citealt{1995PASJ...47..771N, 1997ApJS..110..321B, 2005ApSS.295..107H, 2006ApJ...652..136K, 2008ApJ...678...96B, 2009ApJ...695..276B}; and the project web sites Water Maser Cosmology Project (WMCP)\footnote{\url{https://www.cfa.harvard.edu/~lincoln/demo/HoME/}} and Megamaser Cosmology Project (MCP)\footnote{\url{https://safe.nrao.edu/wiki/bin/view/Main/MegamaserCosmologyProject}}. In comparison, about 500 galaxies have been searched for OH masers, 120 were detected (at up to z\,$=$\,0.265), and about 10 exhibit OH in absorption (see e.g. \citealt{klockner,impellizzeri08}). 

Sources that mase in both molecules are extremely rare. This may be expected, since extragalactic OH and H$_2$O maser species have a quite different pumping mechanism. While OH is radiatively pumped in regions of enhanced density by AGN or star formation photons reprocessed via $\ge45\,$K dust to 35/53\,\textmu{m} infrared (IR), gas phase H$_2$O is collisionally pumped at $\ge400$\,K. 
% M 51 : first and only(?) H2O KM associated with nuclear environment of galaxy, Hagiwara et al. 2001
% H2O MM however just nuclear assoc

Nevertheless, five objects that mase in both species are known: OH and H$_2$O kilomasers (OH~KM, H$_2$O~KM; ${L_\mathrm{OH}, L_\mathrm{H_2O}<10\,L_\sun}$) coexist in the starbursts NGC~253 \citep{1998AJ....115..559F,2004AA...414..117H} and M~82 \citep{1996AA...316..188B,2007MNRAS.380..596A}, an OH~KM and an H$_2$O megamaser (OH~MM, H$_2$O~MM; $L_\mathrm{OH}$, $L_\mathrm{H_2O}>10\,L_\sun$) are found in the Sy~2 NGC~1068 \citep{1996ApJ...462..740G} and radio-quiet Sy~2 AGN NGC~3079 \citep{1995ApJ...446..602B}, whereas Arp~299 is the only OH~MM and H$_2$O~MM object \citep{2007NewAR..51...67T}. No source with an OH~MM and an H$_2$O~KM has yet been found, likely because H$_2$O~KM tend to occur in nearby objects, whereas OH~MM are found up to high redshifts where H$_2$O~KM emission falls below sensitivity limits (see \citealt{2011AA...525A..91T}).

There is no clear general link between OH and H$_2$O maser species in such "dual-species" objects. However, OH maser emission (and OH seen in absorption) may point towards sources that contain denser molecular regions and a generally larger reservoir of H$_2$O; OH is formed from the evaporation and dissociation of grain-bound H$_2$O \citep{annurev.astro.41.011802.094927,2009ApJ...690.1497H}. A larger H$_2$O abundance may favor a H$_2$O maser detection. Furthermore, both H$_2$O~KM, OH~KM and OH~MM have a similar association with, amongst others, star forming regions and nuclear regions with bursts of intense star formation \citep{annurev.astro.41.011802.094927}. We consider that in a H$_2$O maser search, earlier OH detections could be one of the selection criteria that might increase the detection rate.

The search presented here attempts to identify new water maser galaxies, new dual maser species objects and in particular any H$_2$O~KM in an OH~MM galaxy, as well as new AGN disk-maser galaxies suitable for constraining the Hubble constant.

%
%_________________________________________________________________________________________________________________

\section{Source sample}

We considered only sources within 250\,Mpc ($<$\,16\,000\,km\,s$^{-1}$). We first assembled a database of initially over 8000 known water maser detections and non-detections published in the literature, including WMCP and MCP project web site catalogs of published and unpublished maser search results. We then used an automated NASA/IPAC Extragalactic Database (NED) lookup of source coordinates and name aliases to merge duplicates. This produced a final catalog of 4038 unique objects already observed for H$_2$O.  

Next we identified 126 galaxies in the catalog with a low sensitivity H$_2$O non-detection (40\,mJy to 200\,mJy rms) that could be reobserved with a factor 10 higher sensitivity. To identify new sources not yet observed at 22\,GHz, we reviewed recent X-ray and AGN data and literature, including recent extragalactic hydroxyl searches (e.g. \citealt{klockner,impellizzeri08}). We chose objects with {2--10\,keV} X-ray data, high X-ray luminosities (${L_\mathrm{X}  > 10^{40}}$\,W) and large absorbing column densities (${N_\mathrm{H} \gtrsim 10^{22}}$\,cm$^{-2}$) as found in over 90\,\% of water maser galaxies (e.g. \citealt{2006ApJ...652..136K,2006AA...450..933Z,2008ApJ...686L..13G}), and that have a nucleus classified as NLS1, Sy~1.5 to Sy~2.0, LINER or HII region. Some of the sources are also Infrared Astronomical Satellite (IRAS) survey objects with OH masers or OH seen in absorption. 

The final selection of 40 nearby galaxies, mergers, binary AGN and triple AGN systems is shown in Table~\ref{tab:detectionlist} and consists of 17 previous non-detections at $\ge 40$\,mJy rms, 10 objects with OH seen in emission at 1.6\,GHz (3 being previous H$_2$O non-detections), 1 source with strong OH absorption at 6\,GHz (a previous H$_2$O non-detection), and 16 other sources not yet observed for 22\,GHz water masers. Out of this sample, OH maser galaxy ESO~320-G030 had the highest IRAS 100\,\textmu{m} flux density (46\,Jy).

%
%_________________________________________________________________________________________________________________

\section{Observations}

To uncover kilomasers among the source sample a 4\,mJy rms target sensitivity over 24.4 kHz (about 0.31\,km\,s$^{-1}$) was chosen, equivalent to between $0.1\,L_\sun$~rms and $1.0\,L_\sun$~rms in isotropic luminosity. Observations were carried out in four runs during February and March 2013 under project AGBT13A-172 using the National Radio Astronomy Observatory (NRAO\footnote{The National Radio Astronomy Observatory is a facility of the National Science Foundation operated under cooperative agreement by Associated Universities, Inc.}) Robert C. Byrd Green Bank Telescope (GBT). Sessions were scheduled in late winter under good 22\,GHz weather conditions. UTC date and time and zenith opacity $\tau_0$ are listed in Table~\ref{tab:obsdates}. Excellent weather conditions (zenith ${\tau_{0}=0.033}$~to~$0.047$) and low $T_\mathrm{sys}$ (35~K to 75~K) allowed most of the 40 sources to be observed in 16 hours, spending between 12 and 20 minutes on each source and the remainder on calibration.

We used the GBT Spectrometer backend and GBT K-band Focal Plane Array (KFPA) receiver, and selected two KFPA elements near the cryo cooler with the lowest T\subscript{rx} ($<$25~K over 75\,\% of the band). Beams were 33$''$~FWHM with an aperture efficiency $\eta_{ap}$ of 0.66 at 21\,GHz and had a 94.9$''$ beam separation. Observations were in dual-circular polarization and had a 30 second dual-beam "nod" cycle. Both 200~MHz frequency windows of each polarization (i.e. 4 windows in total) were centered on the systemic velocity. The 8192 spectrometer channels were 24.4~kHz wide and covered $V_\mathrm{sys}\pm1250$\,km\,s$^{-1}$.

Pointing and focusing was corrected using strong standard calibrators ($>0.8$\,Jy). GBT dynamic corrections compensated for gravitational and thermal surface deformations and remained stable. The full setup was occasionally verified using brief pointings at megamaser galaxies NGC~1068, NGC~3079 and NGC~5793. Pointing and focusing were repeated every 1.5 to 2 hours and after sunrise, with only small adjustments ($<2.5''$ and $<8$\,mm). Wind speeds were low with little to no cloud cover. Zenith opacity measurements $\tau_0$ at 21\,GHz were supplied by nearby weather stations and remained stable during the observing runs.

\begin{table}
\caption{Observation dates.}\label{tab:obsdates}
\centering
\begin{tabular}{ccccc}
	\hline \hline
	Epoch & Start time (UT)  & Hours & $\tau_0$ \\
	\hline
	 1 & 2013-02-21 20:30 & 3:15 & 0.036--0.037 \\
	 2 & 2013-03-02 03:45 & 3:45 & 0.047 \\
	 ~ & 2013-03-02 07:30 & 1:30 & 0.043 \\
	 ~ & 2013-03-02 09:00 & 1:30 & 0.041 \\
	 3 & 2013-03-03 18:30 & 2:15 & 0.036 \\
	 ~ & 2013-03-03 20:45 & 1:15 & 0.033 \\
	 4 & 2013-03-09 11:15 & 2:15 & 0.036--0.035 \\
	\hline
\end{tabular}
\tablefoot{Start time, duration in hours and $\tau_0$ zenith opacity at 21 GHz of the four K-band observing runs 
in project GBT13A-172.}
\end{table}

\section{Data reduction}

A conservative estimated of the flux scale accuracy is $\le 15$\,\%. All sources were observed in total power mode with $T_\mathrm{cal}$ noise injection to determine $T_\mathrm{sys}$. Although $T_\mathrm{cal}$ has a nominal accuracy of 1\,\%, slow temporal drifts degrade it to 10\,\% -- 15\,\% typical. The initial aperture efficiency $\eta_{ap0}$ calculated by the GBTIDL toolbox was corrected for elevation via the GBT 22.236\,GHz gain-elevation curve\footnote{\url{https://safe.nrao.edu/wiki/bin/view/GB/Observing/GainPerformance}} of April 8, 2008 that is based on the current Zernike model (\textit{FEM~plus~2005WinterV2}) for the GBT adaptive surface. The correction curve is ${\eta_{ap} = \eta_{ap0} * (0.910 + 0.00434*ZD - \num{5.22e-5} *ZD*ZD)}$ where $ZD = 90^\circ\, - \,\mathrm{elevation}$ is the angle off zenith.

Nod scans were processed in GBTIDL. The $\tau_{0}$ and $\eta_{ap}$ gain corrections were applied first. Next, subspectra affected by internal RFI or spectrometer faults were flagged. To improve the SNR, the blank sky reference subspectra (for subtracting the standing wave accross the band) were smoothed with a short 4 to 32 channel boxcar function. Care was taken to avoid introducing artifacts. Third order baselines were removed from individual subspectra. All subspectra and both polarizations were time averaged into a Stokes~I spectrum and a final 3\textsuperscript{rd}-order baseline fit was removed. Sensitivity was calculated over line-free channels without smoothing. % Continuum flux levels are estimates taken from unsmoothed spectra without baseline fits.

For each detection, Gaussian models ${S(\nu) = a*e^{-(\nu-\mu)^{2} / {2\sigma_{\nu}^{2}}}}$ were fitted into the calibrated non-smoothed spectra to estimate line peak $a$~(Jy), center $\mu$~(km\,s$^{-1}$) and FWHM width $w$~(km\,s$^{-1}$). The integrated line profile ${S_\mathrm{int}=\int_{-\infty}^{+\infty} S(\nu)\,d\nu}$ (Jy\,km\,s$^{-1}$) for a single Gaussian $S(\nu)$ equals ${a\,\sigma_{\nu}\sqrt{2\pi}}$. Further noting that ${\sigma_{\nu}=w/2.35482}$, and using the luminosity distance $D_\textrm{L}$ and redshift $z$ of the source, the equivalent isotropic luminosity of
\begin{equation}\label{eq:isolumdef}
\frac{L_\mathrm{iso}}{L_{\sun}}  = 0.023 \times \frac{S_\mathrm{int}}{[\textrm{Jy\,km\,s}^{-1}]} \times \frac{1}{1+z} \times \left( \frac{D_\mathrm{L}}{[\textrm{Mpc}]} \right)^{2}
\end{equation}
can be written for the fitted Gaussians in form of the sum
\begin{equation}\label{eq:isolumgauss}
\frac{L_\mathrm{iso}}{L_{\sun}}  = \frac{0.023 \sqrt{2\pi}}{2.35482} \times \frac{\sum\nolimits_{i} a_{i} \cdot w_{i}}{[\textrm{Jy\,km\,s}^{-1}]} \times \frac{1}{1+z} \times \left( \frac{D_\mathrm{L}}{[\textrm{Mpc}]} \right)^{2}
\end{equation}
over all Gaussian components (see also \citealt{2009ApJ...695..276B}). The factor 0.023 contains unit conversions and the water maser rest frequency (for 1.6\,GHz OH masers the factor is 0.0017). Upper luminosity limits for water maser non-detections are given for single wide Gaussian H$_2$O line with 2.0\,km\,s$^{-1}$ FWHM and a $3\sigma$ peak. The luminosities for detections are based on Eq.~\ref{eq:isolumdef}, with $S_\mathrm{int}$ evaluated directly over the line regions of the spectrum using Simpson's Rule. Spectra of the detections presented here were smoothed with a 3\textsuperscript{rd} order 9-point (2.79\,km\,s$^{-1}$) Savitzky-Golay shape-preserving filter to reduce the noise floor by 70\,\% while maintaining the shape and height of the maser features.

%
%_________________________________________________________________________________________________________________

\section{Results and Discussion}

\begin{figure*}
\centering
\includegraphics[width=0.90\textwidth,angle=-0]{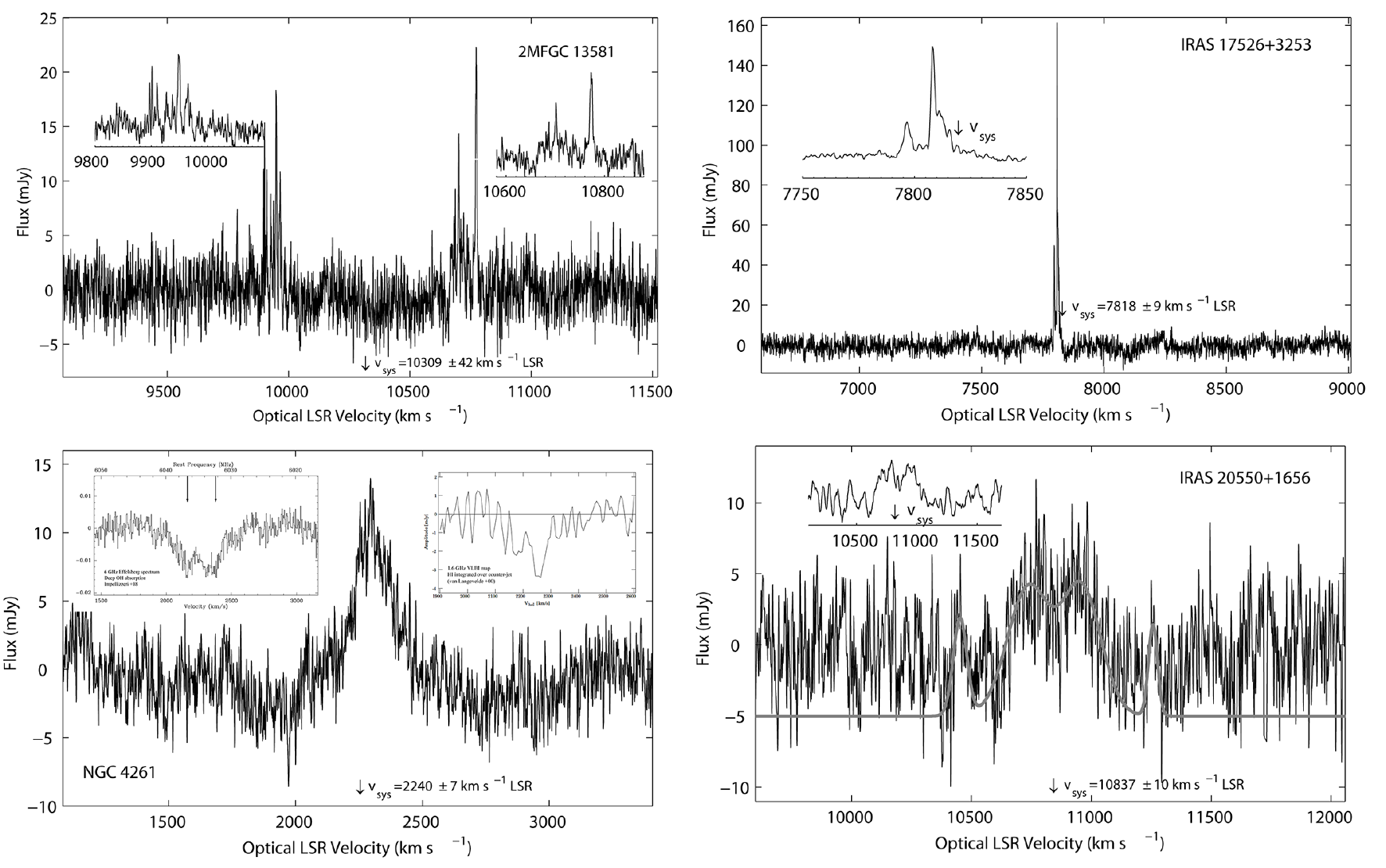}
\caption{Spectra of the four 22\,GHz water megamasers detections. Velocities are in the kinematic Local Standard of Rest frame (LSR) and use the optical convention. Recession velocities, $V_\mathrm{sys}$, are adopted from NED. The channel spacing is 0.3\,km\,s$^{-1}$. The uncertainty of the flux density scale is $\le15\,\%$.
\emph{Top left:} Maser features towards possible Sy~2 galaxy 2MFGC~13581 are typical of masing sub-parsec accretion disks around AGN. 
\emph{Top right:} IRAS~17526+3253 (UGC~11035) is an OH~KM galaxy and has narrow $360\,L_{\sun}$ water masers.
\emph{Bottom left:} NGC~4261 (3C~270; twin-jet, torus) shows deep H\,\Romannum{1}{} absorption at 2260\,km\,s$^{-1}$ (right inset) and deep 6\,GHz OH absorption at 2100--2400\,km\,s$^{-1}$ (left inset) against the counter-jet (insets adopted from \citealt{2000AA...354L..45V,impellizzeri08}). The broad H$_2$O~MM feature is slightly redshifted with respect to the systemic velocity. 
% 2000AA...354L..45V : an HI line at systemic velocity supposedly at 2237 km/s Optical HEL  (not Opt LSR)
%    NGC 4261 : glon=deg2rad(281.8046999);glat=deg2rad(67.3728407); l=glon;b=glat;
%               vLSR = vBSR + 9*cos(l)*cos(b)+12*sin(l)*cos(b)+7*sin(b) = vBSR + 2.65 km/s i.e. LSR-HEL difference not visible in the plot
\emph{Bottom right:} IRAS~20550+1656 (II~Zw~96) is an OH~MM object and has a tentative H$_2$O~MM detection at ${5\sigma}$ after 16-channel Gaussian smoothing (inset). Fits for two main features that are symmetric around $V_\mathrm{sys}$ are overlaid, together with two narrower "suggestive" features that are also symmetric but otherwise similar to the noise peaks even after stronger smoothing.
}
\label{fig:allspec}
\end{figure*}

Of the 40 sources listed in Table~\ref{tab:detectionlist}, 37 were observed for H$_2$O masers with a detection rate of 11\,\%. Three sources have a strong maser detection and one has a tentative detection. The spectra are shown in Figure~\ref{fig:allspec}. Interestingly, three detections are in the set of 11 sources that are also detected in hydroxyl in emission or in deep absorption. No spectra were captured for three sources (NGC~3341, NGC~7130 and ESO~323-G077). The observed sources, sensitivities and isotropic luminosities (H$_2$O and OH), and absorbing column densities are given in Table~\ref{tab:detectionlist}. The luminosity distances, source redshifts and recession velocities, $v_{LSR}$, were adopted from NED. Recession velocities use the optical definition and are in the kinematic Local Standard of Rest (LSR).

Selection by X-ray data did not appear to enhance the H$_2$O maser detection rate. In our sample, 14 active galaxies have a published X-ray absorbing column density. Two of these yielded a H$_2$O~MM detection. One source hosts a heavily obscured AGN ($N_\mathrm{H} > 10^{23}$\,cm$^{-2}$) and the other a Compton-thick AGN ($N_\mathrm{H} \geq 10^{24}$\,cm$^{-2}$). This is consistent with \citet{2013arXiv1309.6515C} who find 96\,\% (45/47) of H$_2$O~MM sources have $N_\mathrm{H} > 10^{23}$\,cm$^{-2}$. However, the same two sources were also selected by OH, resulting in a better detection rate given by the OH selection (3/11) rather than the X-ray selection (2/14).
 % 16 sources have X-ray
 % - 2 are detections and heavily obscured
 % - 2 were not observed
 % s=12 were undetected
 % - 1 was NGC 1167, N_H = 3e21, observed because gas-rich, and has radio jet, but is Compton-thin
 % - 1 was IRAS 17208, N_H = 5.2e21, OH megamaser, but is Compton-thin
 % s=10 
 % Sources with non-detection yet N_H > 10^23 : N833, IRAS 02580, U5881, N4939, U12237  -- five
 %                                subset N_H > 10^24 : U5881, N4939 -- two
 % Sources with non-detection and N_H > 10^23 : N315, N526A, N1167, N1275, U8327, IC4374, IRAS 17208 -- seven

Below we discuss the four water maser detections.

\subsection{2MFGC~13581}

A new H$_2$O disk-MM is found in the optically edge-on galaxy 2MFGC~13581 (${v_\textrm{LSR}=10309\pm42}$\,km\,s$^{-1}$ \citep{2000AAS..142..417H}, ${D_\textrm{L}=145}$\,Mpc, ${z=0.034}$). It is the second-closest of six Seyfert candidates in the Hamburg/SAO survey for emission-line galaxies and is a probable Sy~2 \citep{2000AAS..142..417H}. Masers are detected in two high-velocity groups, symmetrically offset from the systemic velocity by about $\pm$390\,km\,s$^{-1}$. No systemic emission is detected and has a $3\sigma$ upper limit of 6.3\,mJy. The red group peaks at 25\,mJy and forms a forest of $\approx$3\,km\,s$^{-1}$ FWHM lines, similar to the blue group that peaks at 20\,mJy. Each group can be approximated by a wide profile, with the blue one centered at 9940\,km\,s$^{-1}$ (5.3\,mJy peak, 61.5\,km\,s$^{-1}$ FWHM) and the red one at 10724\,km\,s$^{-1}$ (4.4\,mJy peak, 114.4\,km\,s$^{-1}$ FWHM). Their mean, 10332\,km\,s$^{-1}$, lies within $1\sigma$ of the systemic recession velocity. The blue and red groups have luminosities of $120\,L_{\sun}$ and $170\,L_{\sun}$, respectively. % Continuum emission is below 100\,mJy.

The symmetric spectrum suggests that emission most likely originates from a circumnuclear masing disk. The absence of systemic emission may be explained by a warp in the disk that shadows parts of the disk from the X-ray emission by the central engine that is thought to support maser emission \citep{1994ApJ...436L.127N}. The undetected systemic masers may also have a quite low flux density. Alternatively, they may be highly variable like the systemic (disk-)maser emission in Circinus and thus not always detected \citep{2003ApJ...590..162G,2009MNRAS.392.1339M}.

We can estimate the radius of the masing disk annulus in 2MFGC~1358 if we assume Keplerian rotation, a black hole mass of $10^7\,M_{\sun}$ typical of other disk-maser galaxies (e.g. \citealt{2011ApJ...727...20K}), and a fully edge-on disk. The observed orbital velocity of $\pm$390\,km\,s$^{-1}$ then translates into a 0.4\,parsec (0.8\,mas) average disk diameter. Unfortunately, given the low maser flux, prospects for a successful VLBI map and a determination of variability and the secular acceleration of possibly existing weak systemic maser components are poor.

\subsection{IRAS~17526+3253}

Water megamasers in this IR galaxy ({UGC~11035}; ${v_\textrm{LSR}=7818\pm9}$\,km\,s$^{-1}$ \citep{1991rc3..book.....D}, ${D_\textrm{L}=108}$\,Mpc, ${z=0.026}$ (NED), ${L_{IR} \approx \num{7e11} L_{\sun}}$ estimated from IRAS fluxes using the method by \citealt{1986AA...168..237W}) have a relatively high peak flux of ${\approx{170}}$\,mJy. Emission appears in a narrow, slightly blueshifted 40\,km\,s$^{-1}$ window around the systemic velocity. It has three features: two wider profiles at 7797\,km\,s$^{-1}$ (47\,mJy peak, 4.0\,km\,s$^{-1}$ FWHM, $50\,L_{\sun}$) and 7810\,km\,s$^{-1}$ (65\,mJy peak, 10.9\,km\,s$^{-1}$ FWHM, $210\,L_{\sun}$), and a narrow emission line at 7808\,km\,s$^{-1}$ (170 mJy peak, 2.2\,km\,s$^{-1}$ FWHM, $100\,L_{\sun}$). 

Literature lists UGC~11035 as an OH~KM galaxy. An early non-detection (${L_{OH} < 4\,L_\sun}$ at 1.0\,mJy~rms) by \citet{garwood87} was followed by a broad blueshifted feature (3.7\,mJy peak, ${L_{OH} = 6.46\,L_\sun}$) detected near 7450\,km\,s$^{-1}$ by \citealt{1989CRASM.308..287M} but no spectrum was published. A recent observation suffered from RFI and had insufficient sensitivity to confirm the kilomaser \citep{2013ApJ...763....8M}.

The 2MASS image shows two near-infrared components separated by 60$''$ (about 30\,kpc) with an angular size ratio of $\approx 5$. The image suggests UGC~11035 could be a major merger system. The hard X-ray luminosity would then be expected to exceed $10^{43}$\,erg\,s$^{-1}$. This would contribute to enhanced IR luminosity through core emission reprocessed in the dense concentration of gas accumulated during the merger event. Using 60\,\textmu{m}, 100\,\textmu{m}m and S$_{1.4\,\mathrm{GHz}}$ flux points from NED, the $q$ parameter (see e.g. \citealt{2001ApJ...554..803Y}) of this galaxy is 4.74, indicating an IR excess. The $q$ is notably higher than 2.34$\pm$0.01 of the IRAS 2\,Jy sample \citep{2001ApJ...554..803Y} and exceeds 2.57$\pm$0.36 of typical OH~MM galaxies \citep{klockner}, indicating a buried AGN or enhanced star formation. UGC~11035 also has a curiously flat velocity field of 210\,km\,s$^{-1}$ at 9\,kpc from the center, suggesting either a giant irregular galaxy or interaction of two face-on objects. The inner 18\,kpc have a kinematic mass of ${\approx 10^{11}\,M_{\sun}}$ \citep{1992AIPC..254..617A,1994AAS..107...23A}. 

Given limited data on the object we speculate that H$_2$O emission might originate from a small shocked region in an ongoing merger similar to Arp~299. The maser features lack the high velocities associated with jet and nuclear outflow masers (e.g. \citealt{2003ApJ...590..162G}). Their velocity span of 40\,km\,s$^{-1}$ is narrow, but still typical of star formation masers. Assuming a buried AGN gives rise to the IR excess, the systemic masers may also be associated with a slightly inclined, not fully edge-on warped accretion disk. A warp along the line of sight towards the nucleus can compensate for the disk inclination and produce the velocity coherent path lengths necessary for luminous systemic maser emission. Monitoring for secular line accelerations or a VLBI observation are required to rule out a nuclear association.

\subsection{NGC~4261}

The WMCP project lists the giant elliptical galaxy NGC~4261 (${v_\textrm{LSR}=2240\pm7}$\,km\,s$^{-1}$ \citep{2000AJ....119.1645T}, ${D_\textrm{L} = 35.6}$\,Mpc, ${z = 0.0075}$) as a 109\,mJy~rms non-detection observed in 2002. We reached 3.1\,mJy~rms and detect broad emission that is fitted by a single Gaussian of 10.3\,mJy, 154\,km\,s$^{-1}$ FWHM centered on 2302\,km\,s$^{-1}$. Peak emission is redshifted by about +60\,km\,s$^{-1}$ relative to the systemic velocity and has a total isotropic luminosity of $50\,L_{\sun}$. % The continuum flux density is about 0.45\,Jy.

NGC~4261 is a LINER galaxy associated with the low-luminosity FR-\Romannum{1}{} radio source {3C\nobreak\,270} that launches a highly symmetric kpc-scale twin jet. The galaxy is known for its 240\,pc nuclear torus/disk found by the Hubble Space Telescope. The nucleus hosts a \num{4.9e8}\,$M_\sun$ SMBH and has an X-ray absorbing column density $N_\textrm{H}$ of $>\num{5e22}$\,cm$^{-2}$ \citep{2003AA...408..949G}. Recent hard X-ray data show a slightly higher obscuring column density of ${16.4^{21.64}_{13.25} \times 10^{22}}$\,cm$^{-2}$ \citep{gonz09}.

VLBI shows neutral H\,\Romannum{1}{} absorption against the counter-jet, with deepest absorption at 2260\,km\,s$^{-1}$, redwards with respect to the systemic velocity. It has been modeled by atomic gas in a thin disk with an absorbing column density of ${N_\mathrm{H} \approx 10^{21}}$\,cm$^{-2}$ \citep{2000AA...354L..45V}. There is also deep OH absorption at 6\,GHz against a 1.3\,Jy continuum with a $400$\,km\,s$^{-1}$ FWHM around the systemic velocity \citep{impellizzeri08}.

With the presence of a low-luminosity core, an optical dusty torus, molecular absorption and a twin radio jet, NGC~4261 is remarkably similar to the twin-jet LINER NGC~1052. The spectrum of NGC~1052 has a single, broad and slightly redshifted luminous 150\,mJy maser feature. VLBI observations of NGC~1052 found water masers in two regions along the jet axis. Emission has been associated with continuum seed emission from jet blobs being amplified in a X-ray dissociation region located on the inner surface of a torus at a typical (for a J-type shock) temperature of 400\,K \citep{2008ApJ...680..191S}. Given the evidence for abundant H\,\Romannum{1}{} and OH molecules and a high X-ray absorbing $N_\mathrm{H}$ column density towards NGC~4261, combined with the radio jet and the maser profile, a masing torus region with gas infalling at +60\,km\,s$^{-1}$ seems plausible. Existing VLBI datasets map only NGC~4261 continuum emission and do not cover the maser frequency range. A VLBI follow-up is required to determine a masing torus association.

\subsection{IRAS~20550+1656}

We tentatively find broad megamasers in the LIRG/ULIRG irregular galaxy IRAS~20550+1656 (II~Zw~96; ${v_\textrm{LSR}=10837\pm10}$\,km\,s$^{-1}$ \citep{1993AJ....105.1271G}, ${D_\textrm{L}=148}$\,Mpc, ${z=0.036}$). With Arp~299, IRAS~20550+1656 may be the second galaxy that hosts megamasers of both the OH and H$_2$O species. The sensitivity was 5.8\,mJy per 0.3\,km\,s$^{-1}$ channel and 1.5\,mJy after 16-channel Gaussian smoothing. Two 9\,mJy features are seen at $\pm110$\,km\,s$^{-1}$ around the recessional velocity. The combined isotropic luminosity of the blue 10737\,km\,s$^{-1}$ feature (9.0\,mJy peak, 102.8\,km\,s$^{-1}$ FWHM) and the red 10957\,km\,s$^{-1}$ feature (9.1\,mJy peak, 96.4\,km\,s$^{-1}$ FWHM) is relatively high with $600\,L_{\sun}$. Narrower features seem to be symmetrically distributed around the systemic velocity, such as two offset by $-380$\,km\,s$^{-1}$ and $+420$\,km\,s$^{-1}$, but could also be explained as particularly pronounced baseline ripple. These and the broad features are however persistent over different smoothing settings for the "nod" reference spectra, and different baseline fits prior to averaging all subspectra. The broad masers have a post-fit confidence of somewhat better than $5\sigma$. The GBT integration time was 12 minutes and additional time would be needed for a more robust detection especially of the narrower features.

% OH maser spectrum: single wide peak, ~200 km/s FWHM, Staveley-Smith 1987 http://articles.adsabs.harvard.edu//full/1987MNRAS.226..689S/0000696.000.html
% HI emission spectrum: single wide peak, http://www.aanda.org/articles/aa/pdf/2001/10/aa10295.pdf

The object is an ongoing merger or close binary system with a peculiar rotation curve that is known to host an OH~MM with 26\,mJy peak and $83.2\,L_{\sun}$ luminosity \citep{1992AIPC..254..617A,klockner}. H\,\Romannum{1}{} is seen in emission with a FWHM of about $200$\,km\,s$^{-1}$ \citep{2001AA...368...64V}. XMM-Newton observations found a single X-ray source and a high abundance of alpha process elements suggestive of starburst activity, but could not rule out an AGN or AGN-starburst composite. The X-ray core is either very faint or has an ${N_\mathrm{H} > 10^{24}}$\,cm$^{-2}$ \citep{2010AJ....140...63I,2012arXiv1211.1674M}. Starburst activity would be consistent with earlier optical and IR spectroscopy \citep{1997AJ....113.1569G}. VLBI observations of the OH~MM emission have placed it off-center in a merging system. The OH masers trace a 300\,pc region around the second nucleus with a mass of $10^9\,M_{\sun}$ that has some probability of being a heavily obscured AGN \citep{2011MNRAS.416.1267M}. Spitzer observations found starburst activity similar to extranuclear starbursts in NGC~4038/9 and Arp~299. In the latter, three maser regions are associated with both nuclear regions of Arp~299 and an overlap region \citep{2011AA...525A..91T}. The spectrum of the IRAS~20550+1656 water masers is also strongly reminiscent of that of NGC~2146 with its massive star formation kilomasers. The data are quite suggestive that IRAS~20550+1656 masers can be related to starburst and star formation activity. The maser flux is unfortunately rather low for a VLBI follow-up.

\section{Dual maser species}

In literature, there are five known sources (or six, if including uncertain OH masers in UGC~5101) that are known to host both OH and H$_2$O maser species. These "dual-species" objects typically have a complex morphology and the masers are located in unrelated regions. Our two detections in 10 searched OH maser objects increase the number of known dual-species objects to eight.

\begin{figure*}
\centering
\includegraphics[scale=0.72,angle=-0]{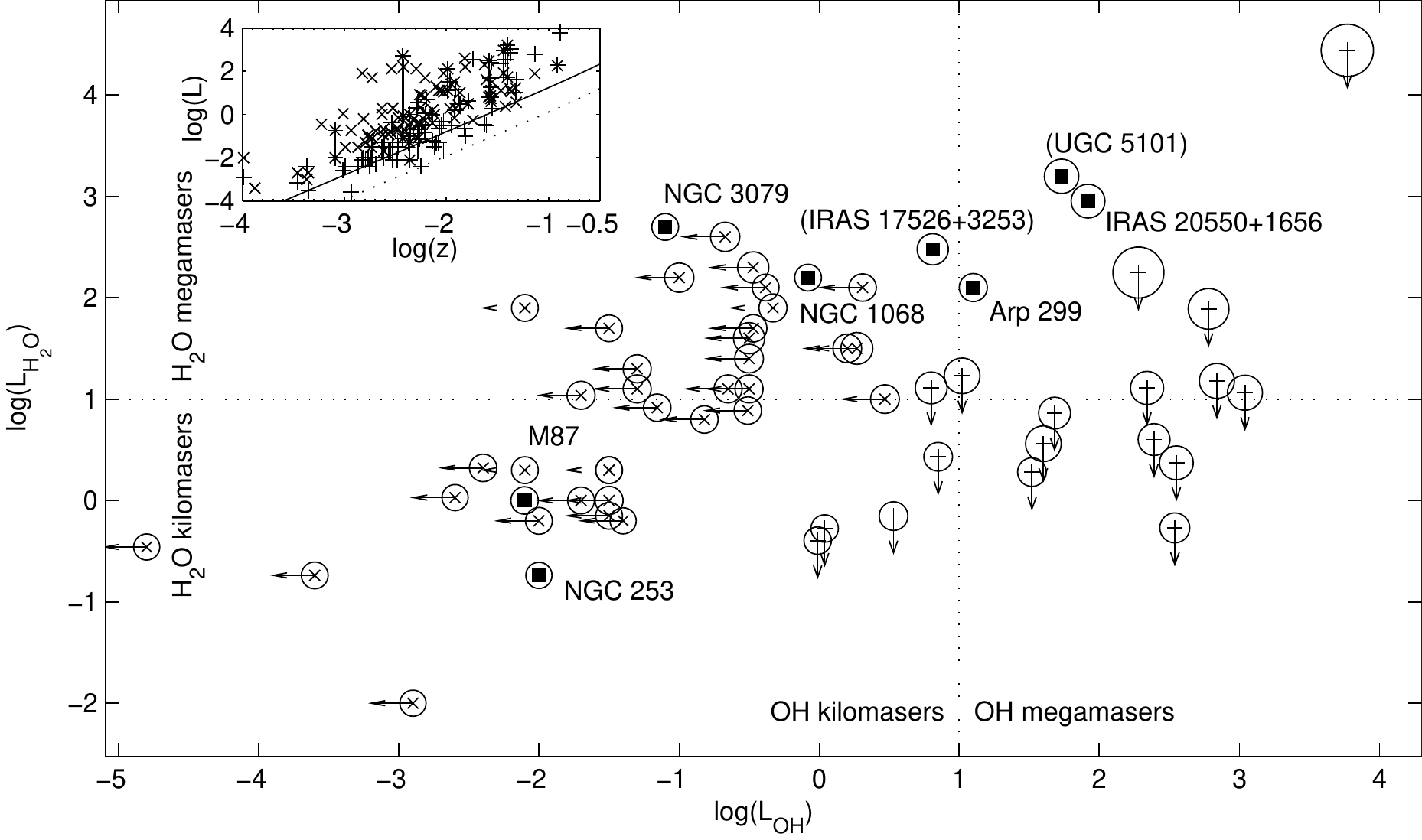}
\caption{Equivalent isotropic luminosities and $3\sigma$ upper limits (arrows) for all 61 sources detected in one or both of the 1.6\,GHz OH and 22\,GHz H$_2$O maser species, with 35 detected in H$_2$O only (x) and 18 in OH only (+). The circle diameters are proportional to the source redshift. The OH~MM sources tend to have a higher redshift. Masers of both species may be found in up to 8 sources (solid squares; source names given), with the caveat that the OH~KM in UGC~5101 and IRAS~17526+3253 reported by \citet{1989CRASM.308..287M} were not detected in later (or earlier) observations. The inset shows OH and H$_2$O luminosities or $3\sigma$ upper limits against redshift for the 61 detected sources and for 34 sources undetected in either species. The H$_2$O and OH detection thresholds (solid and dashed lines) assume 1.0\,mJy rms and a 2.0\,km\,s$^{-1}$ FWHM.}
\label{fig:OHxH2O}
\end{figure*}

\citet{2011AA...525A..91T} report 57 sources searched for both 1.6\,GHz OH and 22\,GHz H$_2$O maser species. Six are detected in neither species, 45 in only one, and six in both species. Tarchi et al. also note a curious lack of H$_2$O~kilomasers in OH~megamaser objects. Adding our sources, and sources common to the 4038-entry H$_2$O database and recent OH searches \citep{1992MNRAS.258..725S,klockner,impellizzeri08,0004-637X-730-1-56}, we find a total of 95 sources searched for both transitions, 34 detected in neither, 35 detected in H$_2$O only, 18 in OH only, and about 8 detected in both species (if the two uncertain OH masers in UGC~5101 and IRAS~17526+3253 are included), for a total of 61 sources detected in at least one of the two maser species. Luminosities, $3\sigma$ limits and redshifts of these 61 detected sources are presented in Figure~\ref{fig:OHxH2O}. Luminosities against source redshifts for all 95 sources are shown in the inset. The upper limits of the non-detections have a median of 0.04~$L_\sun$ for OH and 0.4~$L_\sun$ for H$_2$O. This demonstrates a large fraction of H$_2$O kilomasers ($L<10\,L_{\sun}$) may go undetected. The difference in the two medians is partly due to the frequency proportionality of the isotropic luminosity in Eq.~\ref{eq:isolumdef} -- an at least factor 13 higher sensitivity (lower mJy~rms) is required for detecting H$_2$O~kilomasers at 22\,GHz in a 1.6\,GHz OH~MM object. The probability of detecting a H$_2$O~KM in OH~MM objects is further reduced by the higher average redshift of OH~MM galaxies, with currently known H$_2$O~KM galaxies found at redshifts up to $0.014$ and OH~MM galaxies at redshifts between $0.010$ and $0.264$.

The six reliable dual-species detections correspond to a conditional water maser detection rate of ${P(\mathrm{H_2O}|\mathrm{OH})}=25\,\%$ in the 24 OH~galaxies, and a detection rate of ${P(\mathrm{OH} \cap \mathrm{H_2O}|s_1)=6\,\%}$ for all $N=95$ targets selected for an OH search by some criteria $s_1$ such as high infrared flux density. Over current searches, the average detection rate for OH is about ${P(\mathrm{OH}|s_1)=25\,\%}$ (120 in 500) and about ${P(\mathrm{H_2O}|s_2)=4\,\%}$ for H$_2$O (150 in 4030), where the selection criteria $s_2$ typically include starburst or AGN activity, X-ray luminosity or a high absorbing column density $N_\mathrm{H}$. Accounting for apparently mutually exclusive pumping mechanisms of the OH and H$_2$O species and assuming selections $s_1$ and $s_2$ are independent (although \citealt{2004ASPC..320..199T} find a significantly increased H$_2$O maser detection rate in FIR-luminous sources), the lower bound for the expected probability of two species independently coinciding would be ${P(\mathrm{OH} \cap \mathrm{H_2O}|s_1,s_2) \ge 1\,\%}$, which is on the order of 6\,\% of the $N=95$ dataset. An accurate estimate of the coincidence rate would require a comparison of the selection criteria of each of the 95 sources, a task that is beyond the current scope.

%
%_________________________________________________________________________________________________________________

\section{Conclusions and Summary}

We detected water megamasers in four galaxies: three in OH galaxies, and one in a Sy~2 galaxy with no previous OH search. The OH~MM galaxy IRAS~20550+1656 has a tentative starburst-like H$_2$O~MM detection, the presumed OH~KM galaxy IRAS~17526+3253 has narrow 0.17\,Jy systemic masers, and the 6\,GHz OH absorber NGC~4261 (3C~270) with a twin jet hosts a likely jet-maser. Disk-masers were found towards the probable Sy~2 AGN in 2MFGC~13581. The black hole mass is unknown, but assuming a mass of $10^7\,M_{\sun}$ typical of AGN disk-maser galaxies, the masing disk may be around 0.4\,pc in diameter. Masers in NGC~4261 are particularly interesting as the host galaxy and the maser profile are remarkably similar to H$_2$O~MM galaxy NGC~1052. A VLBI map of NGC~4261 may be able to locate regions of a molecular torus excited by a background jet continuum like in NGC~1052. VLBI imaging of disk-maser galaxy 2MFGC~13581 might produce a precise $M_\mathrm{BH}$ estimate, although the maser flux densities are very low for VLBI. Lack of systemic features in 2MFGC~13581 makes this source less interesting for measuring the angular diameter distance and the Hubble constant. We plan VLBI follow-ups on the more luminous detections in IRAS~17526+3253 and NGC~4261. 

Our detection of water masers in two OH maser galaxies updates the current count of six dual-species objects to eight. The conditional H$_2$O detection rate in OH maser galaxies $P(\mathrm{H_2O}|\mathrm{OH})=25\,\%$ is higher than the average 4\,\% rate achieved in H$_2$O maser searches. We could speculate that OH masers point towards systems with an overall larger molecular reservoir or a larger number of over-densities and shocked regions with favorable conditions for producing detectable luminous H$_2$O maser emission. The dual-species detections in a sample of 95 objects may be explained by a random coincidence of two regions with suitable maser conditions in a single system. However, a more detailed analysis that inspects actual source selection criteria of each of the 95 sources is necessary to determine if dual-species detections are indeed a random coincidence, and if the seemingly enhanced detection rate of H$_2$O masers in extragalactic objects selected by the presence of OH masers is only a product of the small dataset.

%
%_________________________________________________________________________________________________________________

\begin{acknowledgements}
I thank Jim Braatz, Christian Henkel, and Alan Roy for introduction to the GBT, discussion during the proposal, and comments on the manuscript. The author was supported for this research through the International Max Planck Research School (IMPRS) for Astronomy and Astrophysics. This research has made use of the NASA/IPAC Extragalactic Database (NED) which is operated by the Jet Propulsion Laboratory, California Institute of Technology, under contract with the National Aeronautics and Space Administration.
\end{acknowledgements}

%
%_________________________________________________________________________________________________________________

\bibliography{gbt-2013-03-aa}

\begin{thebibliography}{55}
\expandafter\ifx\csname natexlab\endcsname\relax\def\natexlab#1{#1}\fi

\bibitem[{{Andreasian} \& {Alloin}(1994)}]{1994AAS..107...23A}
{Andreasian}, N. \& {Alloin}, D. 1994, \aaps, 107, 23

\bibitem[{{Andreasian}(1992)}]{1992AIPC..254..617A}
{Andreasian}, N.~K. 1992, in American Institute of Physics Conference Series,
  Vol. 254, American Institute of Physics Conference Series, ed. S.~S. {Holt},
  S.~G. {Neff}, \& C.~M. {Urry}, 617--620

\bibitem[{{Argo} {et~al.}(2007){Argo}, {Pedlar}, {Beswick}, \&
  {Muxlow}}]{2007MNRAS.380..596A}
{Argo}, M.~K., {Pedlar}, A., {Beswick}, R.~J., \& {Muxlow}, T.~W.~B. 2007,
  \mnras, 380, 596

\bibitem[{{Baan} \& {Irwin}(1995)}]{1995ApJ...446..602B}
{Baan}, W.~A. \& {Irwin}, J.~A. 1995, \apj, 446, 602

\bibitem[{{Baudry} \& {Brouillet}(1996)}]{1996AA...316..188B}
{Baudry}, A. \& {Brouillet}, N. 1996, \aap, 316, 188

\bibitem[{{Bennert} {et~al.}(2009){Bennert}, {Barvainis}, {Henkel}, \&
  {Antonucci}}]{2009ApJ...695..276B}
{Bennert}, N., {Barvainis}, R., {Henkel}, C., \& {Antonucci}, R. 2009, \apj,
  695, 276

\bibitem[{{Braatz} \& {Gugliucci}(2008)}]{2008ApJ...678...96B}
{Braatz}, J.~A. \& {Gugliucci}, N.~E. 2008, \apj, 678, 96

\bibitem[{{Braatz} {et~al.}(1997){Braatz}, {Wilson}, \&
  {Henkel}}]{1997ApJS..110..321B}
{Braatz}, J.~A., {Wilson}, A.~S., \& {Henkel}, C. 1997, \apjs, 110, 321

\bibitem[{{Castangia} {et~al.}(2013){Castangia}, {Panessa}, {Henkel}, {Kadler},
  \& {Tarchi}}]{2013arXiv1309.6515C}
{Castangia}, P., {Panessa}, F., {Henkel}, C., {Kadler}, M., \& {Tarchi}, A.
  2013, ArXiv e-prints

\bibitem[{{de Vaucouleurs} {et~al.}(1991){de Vaucouleurs}, {de Vaucouleurs},
  {Corwin}, {Buta}, {Paturel}, \& {Fouqu{\'e}}}]{1991rc3..book.....D}
{de Vaucouleurs}, G., {de Vaucouleurs}, A., {Corwin}, Jr., H.~G., {et~al.}
  1991, {Third Reference Catalogue of Bright Galaxies. Volume I: Explanations
  and references. Volume II: Data for galaxies between 0$^{h}$ and 12$^{h}$.
  Volume III: Data for galaxies between 12$^{h}$ and 24$^{h}$.}

\bibitem[{{Frayer} {et~al.}(1998){Frayer}, {Seaquist}, \&
  {Frail}}]{1998AJ....115..559F}
{Frayer}, D.~T., {Seaquist}, E.~R., \& {Frail}, D.~A. 1998, \aj, 115, 559

\bibitem[{{Gallimore} {et~al.}(1996){Gallimore}, {Baum}, {O'Dea}, {Brinks}, \&
  {Pedlar}}]{1996ApJ...462..740G}
{Gallimore}, J.~F., {Baum}, S.~A., {O'Dea}, C.~P., {Brinks}, E., \& {Pedlar},
  A. 1996, \apj, 462, 740

\bibitem[{{Garwood} {et~al.}(1987){Garwood}, {Dickey}, \& {Helou}}]{garwood87}
{Garwood}, R.~W., {Dickey}, J.~M., \& {Helou}, G. 1987, \apj, 322, 88

\bibitem[{{Giovanelli} \& {Haynes}(1993)}]{1993AJ....105.1271G}
{Giovanelli}, R. \& {Haynes}, M.~P. 1993, \aj, 105, 1271

\bibitem[{{Gliozzi} {et~al.}(2003){Gliozzi}, {Sambruna}, \&
  {Brandt}}]{2003AA...408..949G}
{Gliozzi}, M., {Sambruna}, R.~M., \& {Brandt}, W.~N. 2003, \aap, 408, 949

\bibitem[{{Goldader} {et~al.}(1997){Goldader}, {Goldader}, {Joseph}, {Doyon},
  \& {Sanders}}]{1997AJ....113.1569G}
{Goldader}, J.~D., {Goldader}, D.~L., {Joseph}, R.~D., {Doyon}, R., \&
  {Sanders}, D.~B. 1997, \aj, 113, 1569

\bibitem[{{Gonz{\'a}lez-Mart{\'{\i}}n}
  {et~al.}(2009{\natexlab{a}}){Gonz{\'a}lez-Mart{\'{\i}}n}, {Masegosa},
  {M{\'a}rquez}, \& {Guainazzi}}]{gonz09}
{Gonz{\'a}lez-Mart{\'{\i}}n}, O., {Masegosa}, J., {M{\'a}rquez}, I., \&
  {Guainazzi}, M. 2009{\natexlab{a}}, \apj, 704, 1570

\bibitem[{{Gonz{\'a}lez-Mart{\'{\i}}n}
  {et~al.}(2009{\natexlab{b}}){Gonz{\'a}lez-Mart{\'{\i}}n}, {Masegosa},
  {M{\'a}rquez}, \& {Guainazzi}}]{2009ApJ...704.1570G}
{Gonz{\'a}lez-Mart{\'{\i}}n}, O., {Masegosa}, J., {M{\'a}rquez}, I., \&
  {Guainazzi}, M. 2009{\natexlab{b}}, \apj, 704, 1570

\bibitem[{{Greenhill} {et~al.}(2003){Greenhill}, {Booth}, {Ellingsen},
  {Herrnstein}, {Jauncey}, {McCulloch}, {Moran}, {Norris}, {Reynolds}, \&
  {Tzioumis}}]{2003ApJ...590..162G}
{Greenhill}, L.~J., {Booth}, R.~S., {Ellingsen}, S.~P., {et~al.} 2003, \apj,
  590, 162

\bibitem[{{Greenhill} {et~al.}(2008){Greenhill}, {Tilak}, \&
  {Madejski}}]{2008ApJ...686L..13G}
{Greenhill}, L.~J., {Tilak}, A., \& {Madejski}, G. 2008, \apjl, 686, L13

\bibitem[{{Henkel} {et~al.}(2005){Henkel}, {Braatz}, {Tarchi}, {Peck}, {Nagar},
  {Greenhill}, {Wang}, \& {Hagiwara}}]{2005ApSS.295..107H}
{Henkel}, C., {Braatz}, J.~A., {Tarchi}, A., {et~al.} 2005, \apss, 295, 107

\bibitem[{{Henkel} {et~al.}(2004){Henkel}, {Tarchi}, {Menten}, \&
  {Peck}}]{2004AA...414..117H}
{Henkel}, C., {Tarchi}, A., {Menten}, K.~M., \& {Peck}, A.~B. 2004, \aap, 414,
  117

\bibitem[{{Hollenbach} {et~al.}(2009){Hollenbach}, {Kaufman}, {Bergin}, \&
  {Melnick}}]{2009ApJ...690.1497H}
{Hollenbach}, D., {Kaufman}, M.~J., {Bergin}, E.~A., \& {Melnick}, G.~J. 2009,
  \apj, 690, 1497

\bibitem[{{Hopp} {et~al.}(2000){Hopp}, {Engels}, {Green}, {Ugryumov}, {Izotov},
  {Hagen}, {Kniazev}, {Lipovetsky}, {Pustilnik}, {Brosch}, {Masegosa},
  {Martin}, \& {M{\'a}rquez}}]{2000AAS..142..417H}
{Hopp}, U., {Engels}, D., {Green}, R.~F., {et~al.} 2000, \aaps, 142, 417

\bibitem[{{Impellizzeri}(2008)}]{impellizzeri08}
{Impellizzeri}, C. 2008, {Molecular absorption in the cores of AGN: On the
  unified model} (Bonn University Dissertations)

\bibitem[{{Inami} {et~al.}(2010){Inami}, {Armus}, {Surace}, {Mazzarella},
  {Evans}, {Sanders}, {Howell}, {Petric}, {Vavilkin}, {Iwasawa}, {Haan},
  {Murphy}, {Stierwalt}, {Appleton}, {Barnes}, {Bothun}, {Bridge}, {Chan},
  {Charmandaris}, {Frayer}, {Kewley}, {Kim}, {Lord}, {Madore}, {Marshall},
  {Matsuhara}, {Melbourne}, {Rich}, {Schulz}, {Spoon}, {Sturm}, {U},
  {Veilleux}, \& {Xu}}]{2010AJ....140...63I}
{Inami}, H., {Armus}, L., {Surace}, J.~A., {et~al.} 2010, \aj, 140, 63

\bibitem[{{Kl\"ockner}(2004)}]{klockner}
{Kl\"ockner}, H. 2004, {Extragalactic Hydroxyl} (Rijksuniversiteit Groningen)

\bibitem[{{Kl{\"o}ckner} \& {Baan}(2004)}]{2004AA...419..887K}
{Kl{\"o}ckner}, H.-R. \& {Baan}, W.~A. 2004, \aap, 419, 887

\bibitem[{{Kondratko} {et~al.}(2006){Kondratko}, {Greenhill}, \&
  {Moran}}]{2006ApJ...652..136K}
{Kondratko}, P.~T., {Greenhill}, L.~J., \& {Moran}, J.~M. 2006, \apj, 652, 136

\bibitem[{{Kuo} {et~al.}(2011){Kuo}, {Braatz}, {Condon}, {Impellizzeri}, {Lo},
  {Zaw}, {Schenker}, {Henkel}, {Reid}, \& {Greene}}]{2011ApJ...727...20K}
{Kuo}, C.~Y., {Braatz}, J.~A., {Condon}, J.~J., {et~al.} 2011, \apj, 727, 20

\bibitem[{{Lo}(2005)}]{annurev.astro.41.011802.094927}
{Lo}, K.~Y. 2005, Annual Review of Astronomy and Astrophysics, 43, 625

\bibitem[{{Martin} {et~al.}(1989){Martin}, {Bottinelli}, {Gouguenheim}, {Le
  Squeren}, \& {Dennefeld}}]{1989CRASM.308..287M}
{Martin}, J.-M., {Bottinelli}, L., {Gouguenheim}, L., {Le Squeren}, A.-M., \&
  {Dennefeld}, M. 1989, Academie des Sciences Paris Comptes Rendus Serie
  Sciences Mathematiques, 308, 287

\bibitem[{{McBride} \& {Heiles}(2013)}]{2013ApJ...763....8M}
{McBride}, J. \& {Heiles}, C. 2013, \apj, 763, 8

\bibitem[{{McCallum} {et~al.}(2009){McCallum}, {Ellingsen}, {Lovell},
  {Phillips}, \& {Reynolds}}]{2009MNRAS.392.1339M}
{McCallum}, J.~N., {Ellingsen}, S.~P., {Lovell}, J.~E.~J., {Phillips}, C.~J.,
  \& {Reynolds}, J.~E. 2009, \mnras, 392, 1339

\bibitem[{{Migenes} {et~al.}(2011){Migenes}, {Coziol}, {Cooprider},
  {Kl{\"o}ckner}, {Plauchu-Frayn}, {Islas}, \&
  {Ram{\'{\i}}rez-Gardu{\~n}o}}]{2011MNRAS.416.1267M}
{Migenes}, V., {Coziol}, R., {Cooprider}, K., {et~al.} 2011, \mnras, 416, 1267

\bibitem[{{Mudd} {et~al.}(2012){Mudd}, {Mathur}, {Guainazzi}, {Piconcelli},
  {Bianchi}, {Komossa}, {Vignali}, {Lanzuisi}, {Nicastro}, {Fiore}, \&
  {Maiolino}}]{2012arXiv1211.1674M}
{Mudd}, D., {Mathur}, S., {Guainazzi}, M., {et~al.} 2012, \apj~(submitted)

\bibitem[{{Nakai} {et~al.}(1995){Nakai}, {Inoue}, {Miyazawa}, {Miyoshi}, \&
  {Hall}}]{1995PASJ...47..771N}
{Nakai}, N., {Inoue}, M., {Miyazawa}, K., {Miyoshi}, M., \& {Hall}, P. 1995,
  \pasj, 47, 771

\bibitem[{{Neufeld} {et~al.}(1994){Neufeld}, {Maloney}, \&
  {Conger}}]{1994ApJ...436L.127N}
{Neufeld}, D.~A., {Maloney}, P.~R., \& {Conger}, S. 1994, \apjl, 436, L127

\bibitem[{{Noguchi} {et~al.}(2009){Noguchi}, {Terashima}, \&
  {Awaki}}]{2009ApJ...705..454N}
{Noguchi}, K., {Terashima}, Y., \& {Awaki}, H. 2009, \apj, 705, 454

\bibitem[{{Reid} {et~al.}(2013){Reid}, {Braatz}, {Condon}, {Lo}, {Kuo},
  {Impellizzeri}, \& {Henkel}}]{2013ApJ...767..154R}
{Reid}, M.~J., {Braatz}, J.~A., {Condon}, J.~J., {et~al.} 2013, \apj, 767, 154

\bibitem[{{Risaliti} {et~al.}(1999){Risaliti}, {Maiolino}, \&
  {Salvati}}]{1999ApJ...522..157R}
{Risaliti}, G., {Maiolino}, R., \& {Salvati}, M. 1999, \apj, 522, 157

\bibitem[{{Sawada-Satoh} {et~al.}(2008){Sawada-Satoh}, {Kameno}, {Nakamura},
  {Namikawa}, {Shibata}, \& {Inoue}}]{2008ApJ...680..191S}
{Sawada-Satoh}, S., {Kameno}, S., {Nakamura}, K., {et~al.} 2008, \apj, 680, 191

\bibitem[{{Staveley-Smith} {et~al.}(1992){Staveley-Smith}, {Norris}, {Chapman},
  {Allen}, {Whiteoak}, \& {Roy}}]{1992MNRAS.258..725S}
{Staveley-Smith}, L., {Norris}, R.~P., {Chapman}, J.~M., {et~al.} 1992, \mnras,
  258, 725

\bibitem[{{Tan} {et~al.}(2012){Tan}, {Wang}, \& {Zhang}}]{2012ScChG..55.2482T}
{Tan}, Y., {Wang}, J., \& {Zhang}, K. 2012, Science in China G: Physics and
  Astronomy, 55, 2482

\bibitem[{{Tarchi} {et~al.}(2007){Tarchi}, {Castangia}, {Henkel}, \&
  {Menten}}]{2007NewAR..51...67T}
{Tarchi}, A., {Castangia}, P., {Henkel}, C., \& {Menten}, K.~M. 2007, \nar, 51,
  67

\bibitem[{{Tarchi} {et~al.}(2011){Tarchi}, {Castangia}, {Henkel}, {Surcis}, \&
  {Menten}}]{2011AA...525A..91T}
{Tarchi}, A., {Castangia}, P., {Henkel}, C., {Surcis}, G., \& {Menten}, K.~M.
  2011, \aap, 525, A91

\bibitem[{{Tarchi} {et~al.}(2004){Tarchi}, {Henkel}, {Peck}, {Nagar},
  {Moscadelli}, \& {Menten}}]{2004ASPC..320..199T}
{Tarchi}, A., {Henkel}, C., {Peck}, A., {et~al.} 2004, in Astronomical Society
  of the Pacific Conference Series, Vol. 320, The Neutral ISM in Starburst
  Galaxies, ed. S.~{Aalto}, S.~{Huttemeister}, \& A.~{Pedlar}, 199

\bibitem[{{Trager} {et~al.}(2000){Trager}, {Faber}, {Worthey}, \&
  {Gonz{\'a}lez}}]{2000AJ....119.1645T}
{Trager}, S.~C., {Faber}, S.~M., {Worthey}, G., \& {Gonz{\'a}lez}, J.~J. 2000,
  \aj, 119, 1645

\bibitem[{{van Driel} {et~al.}(2001){van Driel}, {Gao}, \&
  {Monnier-Ragaigne}}]{2001AA...368...64V}
{van Driel}, W., {Gao}, Y., \& {Monnier-Ragaigne}, D. 2001, \aap, 368, 64

\bibitem[{{van Langevelde} {et~al.}(2000){van Langevelde}, {Pihlstr{\"o}m},
  {Conway}, {Jaffe}, \& {Schilizzi}}]{2000AA...354L..45V}
{van Langevelde}, H.~J., {Pihlstr{\"o}m}, Y.~M., {Conway}, J.~E., {Jaffe}, W.,
  \& {Schilizzi}, R.~T. 2000, \aap, 354, L45

\bibitem[{{Vasudevan} {et~al.}(2013){Vasudevan}, {Brandt}, {Mushotzky},
  {Winter}, {Baumgartner}, {Shimizu}, {Schneider}, \&
  {Nousek}}]{2013ApJ...763..111V}
{Vasudevan}, R.~V., {Brandt}, W.~N., {Mushotzky}, R.~F., {et~al.} 2013, \apj,
  763, 111

\bibitem[{Willett {et~al.}(2011)Willett, Darling, Spoon, Charmandaris, \&
  Armus}]{0004-637X-730-1-56}
Willett, K.~W., Darling, J., Spoon, H. W.~W., Charmandaris, V., \& Armus, L.
  2011, \apj, 730, 56

\bibitem[{{Wouterloot} \& {Walmsley}(1986)}]{1986AA...168..237W}
{Wouterloot}, J.~G.~A. \& {Walmsley}, C.~M. 1986, \aap, 168, 237

\bibitem[{{Yun} {et~al.}(2001){Yun}, {Reddy}, \&
  {Condon}}]{2001ApJ...554..803Y}
{Yun}, M.~S., {Reddy}, N.~A., \& {Condon}, J.~J. 2001, \apj, 554, 803

\bibitem[{{Zhang} {et~al.}(2006){Zhang}, {Henkel}, {Kadler}, {Greenhill},
  {Nagar}, {Wilson}, \& {Braatz}}]{2006AA...450..933Z}
{Zhang}, J.~S., {Henkel}, C., {Kadler}, M., {et~al.} 2006, \aap, 450, 933

\end{thebibliography}

%
%_________________________________________________________________________________________________________________

% Table with observed sources (name,alias,class,ra,dec, velocity range, rms)
%\input{matlab_observedsources_table_Edited}  % generated by summary_table.m script

\onecolumn
\setcounter{LTchunksize}{50}
\begingroup
\scriptsize
\begin{longtab}
\centering
\begin{longtable}{lclccccccc}
   
   \caption{\normalsize Source sample of the 22 GHz H\subscript{2}O maser search}. \label{tab:detectionlist} \\
   \hline\hline
   
   Source        & Epoch  & \multicolumn{1}{c}{Position}                         & $V_\mathrm{sys}$ & $1\sigma$ $S_{22}$  &  $S_{22}$   & $L_{22}$     & $L_{1.6}$ & $N_\textrm{H}$ & Notes \\
   ~             & ~    & \multicolumn{1}{c}{($\alpha_{2000}$ $\delta_{2000}$)}  & (km\,s$^{-1}$)   & (mJy)               & (mJy)       & ($L_{\sun}$) & ($L_{\sun}$) & (cm$^{-2}$) & ~ \\
   \hline
   \endfirsthead
   
   \caption{continued.} \\
   \hline\hline
   Source        & Epoch  & \multicolumn{1}{c}{Position}                           & $V_\mathrm{sys}$ & $1\sigma$ $S_{22}$  &  $S_{22}$   & $L_{22}$     & $L_{1.6}$ & $N_\textrm{H}$ & Notes \\
   ~             & ~    & \multicolumn{1}{c}{($\alpha_{2000}$ $\delta_{2000}$)}    & (km\,s$^{-1}$)   & (mJy)               & (mJy)       & ($L_{\sun}$) & ($L_{\sun}$) & (cm$^{-2}$) & ~ \\
   \hline
   \endhead
   
   \hline
   \endfoot
   
01 IRAS 00160-0719 $^{(a)}$          & 3   & 00:18:35.9 -07:02:56.0 & $  5396$ & $2.9$  &  $<8.7$ & $<2.06$  & ~  & ~  & Sy2+HII\\*
02 ESO 350-IG038 $^{(a)}$            & 3   & 00:36:52.7 -33:33:17.0 & $  6175$ & $4.6$  & $<13.8$ & $<4.48$  & ~  & ~  & pair, LIRG, HII/SB\\*
03 IRAS 00494-3056 $^{(a)}$          & 3   & 00:51:51.8 -30:40:00.0 & $ 15529$ & $5.3$  & $<15.9$ & $<34.86$  & ~  & ~  & SB, ULIRG\\*
04 NGC 0315 $^{(a)}$                 & 3   & 00:57:48.9 +30:21:09.0 & $  4942$ & $3.0$  &  $<9.0$ & $<1.54$  & ~  & $\num{5.0e+022}$  & LINER(?), jet\\*
05 NGC 0526A                         & 1   & 01:23:54.4 -35:03:56.0 & $  5725$ & $4.3$  & $<12.9$ & $<3.62$  & ~  & $\num{2.0e+022}$  & Sy1.9\\*
06 \textit{IRAS 01418+1651}          & 3   & 01:44:30.5 +17:06:05.0 & $  8225$ & $2.9$  &  $<8.7$ & $<5.11$  &  {\textit{$245.0$}}  & ~  & Sy2, LIRG\\*
07 NGC 0833                          & 1   & 02:09:20.8 -10:08:59.0 & $  3864$ & $2.9$  &  $<8.7$ & $<1.05$  & ~  & $\num{2.7e+023}$  & Sy2/LINER\\*
08 NGC 0925 $^{(a)}$                 & 3   & 02:27:16.9 +33:34:45.0 & $   553$ & $2.7$  &  $<8.1$ & $<0.03$  & ~  & ~  & LLAGN\\*
09 IC 1858 $^{(a)}$                  & 1   & 02:49:08.4 -31:17:22.0 & $  6070$ & $3.5$  & $<10.5$ & $<3.41$  & ~  & ~  & LINER(?)\\*
10 IRAS 02580-1136                   & 1   & 03:00:30.6 -11:24:57.0 & $  8962$ & $3.5$  & $<10.5$ & $<8.37$  & ~  & $\num{5.6e+023}$  & Sy2+HII, merger\\
11 NGC 1167 $^{(a)}$                 & 1   & 03:01:42.4 +35:12:21.0 & $  4945$ & $2.7$  &  $<8.1$ & $<1.70$  & ~  & $\num{3.0e+021}$  & Sy2/LINER, jet\\
12 NGC 1275 $^{(a,b)}$               & 3   & 03:19:48.1 +41:30:42.0 & $  5264$ & $26.4$ & $<79.2$ & $<17.73$  & ~  & $\num{1.5e+022}$  & Sy1.5\\
13 IRAS F04023-1638 $^{(a)}$         & 1   & 04:04:40.6 -16:30:14.0 & $  8643$ & $3.1$  &  $<9.3$ & $<6.37$  & ~  & ~  & n/a\\
14 \textit{IRAS 04332+0209}          & 1   & 04:35:48.4 +02:15:29.0 & $  3590$ & $2.7$  &  $<8.1$ & $<0.93$  &  {\textit{$1.9$}}  & ~  & HII\\
15 MCG+02-21-013                     & 2   & 08:04:46.4 +10:46:36.0 & $ 10323$ & $4.1$  & $<12.3$ & $<12.93$  & ~  & ~  & pair, Sy2+Sy1\\
16 NGC 2974                          & 2   & 09:42:33.3 -03:41:57.0 & $  1919$ & $3.0$  &  $<9.0$ & $<0.27$  & ~  & ~  & Sy2, SF\\
17 \textit{ESO 374-IG032} $^{(a)}$   & 2   & 10:06:05.1 -33:53:17.0 & $ 10223$ & $5.3$  & $<15.9$ & $<16.49$  &  {\textit{$724.0$}}  & ~  & LIRG, merger, pair\\
18 NGC 3341                          & --  & 10:42:31.5 +05:02:38.0 & $ 8196$  & --     & --      & --        & ~ & ~ & Sy2+LINER+comp \\
19 UGC 05881                         & 2   & 10:46:42.5 +25:55:54.0 & $  6173$ & $2.9$  &  $<8.7$ & $<3.40$  & ~  & $\num{2.5e+024}$  & LINER\\
20 NGC 3563B $^{(a)}$                & 2   & 11:11:25.2 +26:57:48.9 & $ 10774$ & $3.0$  &  $<9.0$ & $<10.35$  & ~  & ~  & jet, FSRS\\
21 NGC 3655 $^{(a)}$                 & 2   & 11:22:54.6 +16:35:24.1 & $  1473$ & $2.8$  &  $<8.4$ & $<0.39$  & ~  & ~  & HII\\
22 \textit{ESO320-G030} $^{(a)}$     & 2   & 11:53:11.7 -39:07:49.0 & $  3232$ & $6.9$  & $<20.7$ & $<2.41$  &  {\textit{$33.0$}}  & ~  & LIRG, SB, AGN\\
23 \textbf{\textit{NGC 4261}} $^{(a,c)}$ & 2   & \textbf{\textit{12:19:23.2 +05:49:31.0}}& $2238$ & $3.1$  & $\textbf{\textit{10.3}}$ & $\textbf{\textit{50}}$  &  abs. & $\num{1.6e+023}$  & LINER, twin-jet\\
24 NGC 4939 $^{(a)}$                 & 4   & 13:04:14.4 -10:20:23.0 & $  3110$ & $12.8$ & $<38.4$ & $<2.83$  & ~  & $>\num{1.0e+025}$  & Sy2\\
25 ESO 323-G077                      & --  & 13:06:26.1 -40:24:53.0 & $  4501$ & --     & --      & --        & ~ & $\num{5.5e+023}$ & Sy1.2/LINER \\
26 UGC 8327                          & 2   & 13:15:15.6 +44:24:26.0 & $ 11002$ & $2.7$  &  $<8.1$ & $<9.55$  & ~  & $\num{5.2e+022}$  & Sy2+Sy2\\
27 IC 4374 $^{(a)}$                  & 2   & 14:07:29.8 -27:01:04.3 & $  6535$ & $4.0$  & $<12.0$ & $<5.16$  & ~  & $>\num{8.0e+021}$  & FSRS, FRII\\
28 NGC 5806                          & 2   & 15:00:00.4 +01:53:29.0 & $  1359$ & $3.4$  & $<10.2$ & $<0.32$  & ~  & ~  & Sy2\\
29 \textit{IRAS 15065-1107} $^{(a)}$ & 2   & 15:08:49.1 -11:12:28.0 & $  2105$ & $3.6$  & $<10.8$ & $<0.53$  &  $1.1$ & ~  & pair\\
30 \textit{IRAS 15179+3956}          & 4   & 15:19:47.1 +39:45:38.0 & $ 14261$ & $2.9$  &  $<8.7$ & $<16.92$  &  $10.5$ & ~  & 2core, HII\\
31 Ark 481 $^{(a)}$                  & 2   & 15:39:05.2 +05:34:16.6 & $  7781$ & $3.3$  &  $<9.9$ & $<5.82$  & ~  & ~  & n/a\\
32 \textit{IRAS 16399-0937}          & 2   & 16:42:40.2 -09:43:14.0 & $  8098$ & $3.9$  & $<11.7$ & $<7.25$  &  $47.0$ & ~  & Sy2+HII/LINER\\
33 \textbf{2MFGC13581}               & 2+4 & \textbf{16:58:15.5 +39:23:29.0} & $ 10290$ &  $2.1$  & $\textbf{25.5}$ & $\textbf{290}$  & ~  & ~  & Sy2\\
34 \textit{IRAS 17208-0014}          & 2+4 & 17:23:21.9 -00:17:01.0 & $ 12834$ & $2.5$  &  $<7.5$ & $<11.54$  &  $1090.0$ & $\num{5.2e+021}$  & HII, LINER, SB\\
35 \textbf{\textit{IRAS 17526+3253}} & 4   & \textbf{\textit{17:54:29.4 +32:53:14.0}} & $7798$ & $5.3$  & $\textbf{\textit{170}}$ & $\textbf{\textit{360}}$  & $6.5$ & ~  & ULIRG,LINER(?)\\
36 UGC 11185NED01                    & 2   & 18:16:09.4 +42:39:23.0 & $ 12426$ & $3.0$  &  $<9.0$ & $<12.81$  & ~  & ~  & Sy2+Sy2\\
37 \textbf{\textit{IRAS 20550+1656}} & 4   & \textbf{\textit{20:57:23.9 +17:07:39.0}} & $ 10822$ & $5.8^{(d)}$  & $\textbf{\textit{9.1}}$ & $\textbf{\textit{600}}$  &  $83.2$  & $>\num{1.0e+024}$  & LIRG, HII, pair\\
38 NGC 7130                          & --  & 21:48:19.5 -34:57:04.0 & $  4842$ & --     & --      & --     & ~  & $> \num{6.0e+023}$ & Sy2/LINER/HII \\
39 UGC 12237                         & 3   & 22:54:19.7 +11:46:57.0 & $  8476$ & $3.1$  &  $<9.3$ & $<7.01$  & ~  & $\num{1.9e+023}$  & Sy2/SB, SNIa \\
40 NGC 7742                          & 3   & 23:44:15.8 +10:46:01.0 & $  1663$ & $1.8$  &  $<5.4$ & $<0.13$  & ~  & ~  & Sy2/LINER,SB,HII\\
\end{longtable}
\tablefoot{\normalsize
Epochs refer to observing dates (see Table \ref{tab:obsdates}). The three sources without epochs were not observed. 
Optical heliocentric recession velocities ($V_\mathrm{sys}$) are adopted from NED. The maser search covered $V_\mathrm{sys} \pm 1250$\,km\,s$^{-1}$. 
The flux scale is accurate to better than 15\,\%. 
Sensitivity ($1\sigma\,S_{22}$) is the rms over the line-free 75\,\% of the 200\,MHz band and is derived from 24.4\,kHz channels without smoothing. 
The peak flux ($S_{22}$) and total equivalent isotropic luminosity ($L_{22}$) are stated for detections (sources in bold; 23, 33, 35, 37). 
For non-detections, $3\sigma$ upper limits on $S_{22}$ and $L_{22}$ are given, with $L_{22}$ based on $3 \cdot 1\sigma\,S_{22}$ and a 2.0\,km\,s$^{-1}$ FWHM Gaussian.
Sources with hydroxyl (OH) masers are indicated in italic and the OH maser luminosities ($L_{1.6}$) are adopted from \citet{1992MNRAS.258..725S,klockner,impellizzeri08,0004-637X-730-1-56}.
The 2--10\,keV X-ray absorbing column densities ($N_\textrm{H}$) are adopted from literature (\citealt{1999ApJ...522..157R,2009ApJ...704.1570G,2009ApJ...705..454N,2012ScChG..55.2482T,2013ApJ...763..111V}). 
The last column shows the object classification adopted from NED and literature. \\
Remarks: 
(a) Source has a previous published or unpublished H$_2$O non-detection (20\,mJy to 200\,mJy rms), with references on the WMCP or MCP project pages.
(b) The velocity coverage on NGC~1275 was 3000 to 14000\,km\,s$^{-1}$ (800\,MHz) to include the infalling object at +3000\,km\,s$^{-1}$.
(c) NGC~4261 exhibits H\,\Romannum{1}{} in absorption at 1.4\,GHz and OH in deep absorption at 6\,GHz, but has not been searched for OH at 1.6\,GHz \citep{2000AA...354L..45V,impellizzeri08}.
(d) Sensitivity on IRAS~20550+1656 was 5.8\,mJy without smoothing and 1.5\,mJy rms after 16-channel Gaussian smoothing.
}
\end{longtab}
\endgroup

%
%_________________________________________________________________________________________________________________

\onecolumn

\twocolumn

%
%_________________________________________________________________________________________________________________

\end{document}